\documentclass{nature}

\usepackage{amssymb}
\usepackage{graphicx}
\usepackage{color}
\usepackage[normalem]{ulem}
\usepackage{soul}

\usepackage{subfig}

\usepackage{floatrow}
\DeclareFloatFont{tiny}{\scriptsize}
\title{Direct detection of a break in the teraelectronvolt
cosmic-ray spectrum of electrons and positrons}

\newcounter{firstbib}

\begin{document}

\maketitle

\author{DAMPE Collaboration:
G.~Ambrosi$^{3}$, Q.~An$^{4,5}$, R.~Asfandiyarov$^{6}$, P.~Azzarello$^{6}$, P.~Bernardini$^{7,8}$, B.~Bertucci$^{9,3}$, M.~S.~Cai$^{1,2}$,
J.~Chang$^{1,2}$, D.~Y.~Chen$^{1,10}$, H.~F.~Chen$^{4,5}$
% ...
}
\begin{affiliations}
\small
\item{Key Laboratory of Dark Matter and Space Astronomy, Purple Mountain Observatory, Chinese Academy of Sciences, Nanjing 210008, China}
\item{School of Astronomy and Space Science, University of Science and Technology of China, Hefei 230026, China}
\item{Istituto Nazionale di Fisica Nucleare (INFN) - Sezione di Perugia, I-06123 Perugia, Italy}
\item{State Key Laboratory of Particle Detection and Electronics, University of Science and Technology of China, Hefei 230026, China}
\item{Department of Modern Physics, University of Science and Technology of China, Hefei 230026, China}
\item{Department of Nuclear and Particle Physics, University of Geneva, CH-1211, Switzerland}
\item{Dipartimento di Matematica e Fisica ¡°E. De Giorgi¡±, Universit\`a del Salento, I-73100, Lecce, Italy}
\item{Istituto Nazionale di Fisica Nucleare (INFN) - Sezione di Lecce, I-73100, Lecce, Italy}
\item{Dipartimento di Fisica e Geologia, Universit\`a degli Studi di Perugia, I-06123 Perugia, Italy}
\item{University of Chinese Academy of Sciences, Yuquan Road 19, Beijing 100049, China}
\item{Institute of Modern Physics, Chinese Academy of Sciences, Nanchang Road 509, Lanzhou 730000, China}
\item{National Space Science Center, Chinese Academy of Sciences, Nanertiao 1, Zhongguancun, Haidian district, Beijing 100190, China}
\item{Institute of High Energy Physics, Chinese Academy of Sciences, YuquanLu 19B, Beijing 100049, China}
\item{Istituto Nazionale di Fisica Nucleare (INFN) - Sezione di Bari, I-70125, Bari, Italy}
\item{ASI Space Science Data Center (SSDC), I-00133 Roma, Italy}
\item{Dipartimento di Fisica ``M.~Merlin'' dell'Univerisit\`a e del Politecnico di Bari, I-70126, Bari, Italy}
\item{Department of Physics and Laboratory for Space Research, the University of Hong Kong, Pok Fu Lam, Hong Kong SAR, China}
\end{affiliations}

\hfill

\begin{abstract}
High energy cosmic ray electrons plus positrons (CREs), which lose energy quickly during their propagation, provide an ideal probe of Galactic high-energy processes\cite{Meyer1969,Strong2007,Fan2010,Shen1970,Aharonian1995,Kobayashi2004,Moskalenko1998} and may enable the observation
of phenomena such as dark-matter particle annihilation or decay\cite{Turner1990,Bertone2005,Feng2010}.
The CRE spectrum has been directly measured up to $\sim 2$ TeV in previous balloon- or space-borne experiments\cite{Nishimura1980,Chang2008b,Abdo2009,Aguilar2014,Abdollahi2017,Adriani2011}, and indirectly up to $\sim 5$ TeV by ground-based Cherenkov $\gamma$-ray telescope arrays\cite{Aharonian2008,Aharonian2009}. Evidence for a spectral break in the TeV
energy range has been provided by indirect measurements of H.E.S.S.\cite{Aharonian2008,Aharonian2009}, although the results were qualified by sizeable systematic uncertainties.
Here we report a direct measurement of CREs in the energy range $25~{\rm GeV}-4.6~{\rm TeV}$ by the DArk Matter Particle Explorer (DAMPE)\cite{Chang2017} with unprecedentedly high energy resolution and low background.
The majority of the spectrum can be properly fitted by a smoothly broken power-law model rather than a single power-law model.
The direct detection of a spectral break at $E \sim0.9$ TeV confirms the evidence found by H.E.S.S., clarifies the behavior of the CRE spectrum at energies above 1 TeV and sheds light on the physical origin of the sub-TeV CREs.
\end{abstract}
%

The DArk Matter Particle Explorer (DAMPE; also known as ``Wukong'' in China), which was launched into a sun-synchronous orbit at an altitude of  $\sim$500 km on December 17, 2015, is a high energy particle detector optimized for studies of CREs and $\gamma$-rays up to $\sim 10$ TeV. The DAMPE instrument, from top to bottom, consists of a Plastic Scintillator Detector (PSD), a Silicon-Tungsten tracKer-converter detector (STK), a BGO imaging calorimeter, and a NeUtron Detector (NUD)\cite{Chang2017}.
The PSD measures the charge of incident particles with a high nuclear resolution up to $Z=28$, and
aids in the discrimination between photons and charged particles. The STK measures the charge and trajectory of charged particles, and reconstructs the direction of $\gamma$-rays converting into $e^+e^-$ pairs. The BGO calorimeter\cite{ZhangZY2015}, with a total depth of $\sim32$ radiation lengths and $\sim 1.6$ nuclear interaction lengths, measures the energy of incident particles and provides efficient CRE identification. The NUD further improves the electron/proton discrimination at TeV energies\cite{Chang2017}. With these four sub-detectors combined, DAMPE has achieved a high rejection power of the hadronic cosmic ray background, a large effective acceptance, and much improved energy resolution for CRE measurements\cite{Chang2017}.
In 2014-2015 the Engineering Qualification Model (see the Methods) was extensively tested using test beams at the European Organization for Nuclear Research (CERN). The beam test data demonstrated the excellent energy resolution for electrons and $\gamma$-rays (better than 1.2\% for energies $>100$ GeV \cite{YueChuan,ZhangZY2016}), and verified the electron/proton discrimination capabilities\cite{Chang2017} consistent with simulation results.

The cosmic ray proton-to-electron flux ratio increases from $\sim 300$ at 100 GeV to $\sim 800$ at 1 TeV. A robust electron/proton discrimination and an accurate estimate of the residual proton background are therefore crucial for reliable measurement of CRE spectrum. As the major instrument onboard DAMPE, the BGO calorimeter ensures a well-contained development of electromagnetic showers in the energy range of interest. The electron/proton discrimination method relies on an image-based pattern recognition, as adopted in the ATIC experiment\cite{Chang2008a}. It exploits the topological differences of the shower shape between hadronic and electromagnetic particles in the BGO calorimeter. This method, together with the event pre-selection procedure, is found to be able to reject $>99.99\%$ of the protons while keeping $90\%$ of the electrons and positrons. The details of the electron identification are presented in the Methods (for example, in Extended Data Figure~1 we show the consistency of the electron/proton discrimination between the flight data and the Monte Carlo simulations). Figure~1 illustrates the discrimination power of DAMPE between electrons and protons with deposit energies of $500-1000$ GeV, using the BGO images only.

The results reported in this work are based on data recorded between December 27, 2015 and June 8, 2017. Data collected while the satellite passing the South Atlantic Anomaly has been excluded in the analysis. During these $\sim530$ days of operation, DAMPE has recorded more than 2.8 billion cosmic ray events, including $\sim$1.5 million CREs above 25 GeV. Figure~2 shows the corresponding CRE spectrum measured from the DAMPE data (see Table 1 for more details), compared with previously published results from the space-borne experiments AMS-02\cite{Aguilar2014} and Fermi-LAT\cite{Abdollahi2017}, as well as the ground-based H.E.S.S. experiment\cite{Aharonian2008,Aharonian2009}. The contamination of the proton background for DAMPE is estimated to be less than 3\% in the energy range of 50 GeV$-1$ TeV (see Table 1).
The systematic uncertainties of the flux measurement have been evaluated, with dominant contributions from the background subtraction and the instrumental effective acceptance (the product of the fiducial instrumental acceptance and the particle selection efficiency). More details on the systematic uncertainties can be found in the Methods.

A spectral hardening at $\sim50$ GeV is shown in our data, in agreement with that of AMS-02 and Fermi-LAT.
The data in the energy range of $55~{\rm GeV}-2.63~{\rm TeV}$ strongly prefer a smoothly broken power-law model (the fit yields a $\chi^{2}=23.3$ for 18 degrees of freedom) to a single power-law model (which yields a $\chi^{2}=70.2$ for 20 degrees of freedom).
Our direct detection of a spectral break at $E \sim0.9$ TeV, with the spectral index changing from $\gamma_{1}\sim3.1$ to $\gamma_{2}\sim3.9$ (see the Methods for details), confirms
the previous evidence found by the ground-based indirect measurement of H.E.S.S.\cite{Aharonian2008,Aharonian2009}. The AMS-02 data also predict a TeV spectral softening with the so-called minimal model\cite{Accardo2014}. Our results are consistent with the latest CRE spectra measured by Fermi-LAT\cite{Abdollahi2017} in a wide energy range, although the $\sim$TeV break has not been detected by Fermi-LAT, possibly due to higher particle background contamination and/or lower instrumental energy resolution.
We note that the CRE flux measured by DAMPE is overall higher than the one reported by AMS-02 for energies $\gtrsim70$ GeV. The difference might be partially due to the uncertainty in the absolute energy scale, which would coherently shift the CRE spectrum up or down. With increased statistics and improved understanding of the detector's performance, more consistent measurements among different experiments might be achieved in the near future.

Different from the H.E.S.S. data which is systematic uncertainties dominated, the DAMPE CRE spectrum shown in Figure~2 is statistical uncertainties dominated for energies above $\sim 380$ GeV. DAMPE is designed to operate for at least three years, and likely will be extended to a longer lifetime given the current instrument status. Further increased statistics will allow more precise measurement of the CRE spectrum up to higher energies $\sim$10 TeV and crucially test whether there is any edge-like feature that may be generated by dark matter annihilation/decay or nearby pulsars\cite{Shen1970,Kobayashi2004,Hooper2017}. The precise measurement of the CRE spectrum by DAMPE can narrow down the parameter space of models such as nearby pulsars, supernova remnants, and/or candidates of particle dark matter considerably \cite{Fan2010,Blasi2009,Hooper2009,Bergstrom2009,Fermi2017} in order to  account for the ``positron anomaly" \cite{Adriani2009,Accardo2014}.  The parameters include, for example, the spectral cutoff energy of the electrons accelerated by nearby pulsars or supernova remnants, or the rest mass and the annihilation cross section (or alternatively the lifetime) of a dark matter particle. Together with data from the cosmic microwave background or $\gamma$-rays, these improved constraints on the model parameters obtained by DAMPE may ultimately clarify the connection between the positron anomaly and the annihilation or decay of particle dark matter.

\newpage

\begin{table}
\begin{center}
\title{}Table 1. {\bf The CRE flux in units of $({\rm m^{-2}s^{-1}sr^{-1}GeV^{-1}})$, together with 1$\sigma$ statistical and systematic errors.}\\
\begin{tabular}{ r@{~--~}l r@{~$\pm$~}l r@{~$\pm$~}l c r@{~$\pm$~}l r@{~$\pm$~}c@{~$\pm$~}l }
\hline
 \multicolumn{2}{c}{Energy range (GeV) }  &  \multicolumn{2}{c}{$\langle E\rangle$ (GeV) } & \multicolumn{2}{c}{Acceptance (m$^2\times$sr)} & Counts &  \multicolumn{2}{c}{Bkg.~fraction} & \multicolumn{3}{c}{$\Phi$(e$^+$+e$^-$)$\pm$$\sigma_{\mathrm{stat}}$$\pm$$\sigma_{\mathrm{sys}}$}\\
\hline
      24.0 &     27.5 &     25.7 &      0.3 &    0.256 &    0.007 &     377469 & ( 2.6 &  0.3)\% & (1.16 & 0.00 & 0.03)$\times$10$^{-2}$  \\ 
      27.5 &     31.6 &     29.5 &      0.4 &    0.259 &    0.007 &     279458 & ( 2.5 &  0.3)\% & (7.38 & 0.02 & 0.19)$\times$10$^{-3}$  \\ 
      31.6 &     36.3 &     33.9 &      0.4 &    0.261 &    0.007 &     208809 & ( 2.4 &  0.2)\% & (4.76 & 0.02 & 0.13)$\times$10$^{-3}$  \\ 
      36.3 &     41.7 &     38.9 &      0.5 &    0.264 &    0.007 &     156489 & ( 2.4 &  0.2)\% & (3.08 & 0.01 & 0.08)$\times$10$^{-3}$  \\ 
      41.7 &     47.9 &     44.6 &      0.6 &    0.266 &    0.007 &     117246 & ( 2.3 &  0.2)\% & (2.00 & 0.01 & 0.05)$\times$10$^{-3}$  \\ 
      47.9 &     55.0 &     51.2 &      0.6 &    0.269 &    0.007 &      87259 & ( 2.3 &  0.2)\% & (1.28 & 0.01 & 0.03)$\times$10$^{-3}$  \\ 
      55.0 &     63.1 &     58.8 &      0.7 &    0.272 &    0.007 &      65860 & ( 2.2 &  0.2)\% & (8.32 & 0.04 & 0.21)$\times$10$^{-4}$  \\ 
      63.1 &     72.4 &     67.6 &      0.8 &    0.275 &    0.007 &      49600 & ( 2.1 &  0.2)\% & (5.42 & 0.03 & 0.13)$\times$10$^{-4}$  \\ 
      72.4 &     83.2 &     77.6 &      1.0 &    0.277 &    0.007 &      37522 & ( 2.1 &  0.2)\% & (3.54 & 0.02 & 0.09)$\times$10$^{-4}$  \\ 
      83.2 &     95.5 &     89.1 &      1.1 &    0.279 &    0.007 &      28325 & ( 2.1 &  0.1)\% & (2.31 & 0.01 & 0.06)$\times$10$^{-4}$  \\ 
      95.5 &    109.7 &    102.2 &      1.3 &    0.283 &    0.007 &      21644 & ( 2.0 &  0.1)\% & (1.52 & 0.01 & 0.04)$\times$10$^{-4}$  \\ 
     109.7 &    125.9 &    117.4 &      1.5 &    0.282 &    0.007 &      16319 & ( 2.0 &  0.1)\% & (1.00 & 0.01 & 0.02)$\times$10$^{-4}$  \\ 
     125.9 &    144.5 &    134.8 &      1.7 &    0.286 &    0.007 &      12337 & ( 2.0 &  0.1)\% & (6.49 & 0.06 & 0.16)$\times$10$^{-5}$  \\ 
     144.5 &    166.0 &    154.8 &      1.9 &    0.287 &    0.007 &       9079 & ( 2.0 &  0.1)\% & (4.14 & 0.04 & 0.10)$\times$10$^{-5}$  \\ 
     166.0 &    190.6 &    177.7 &      2.2 &    0.288 &    0.007 &       7007 & ( 1.9 &  0.1)\% & (2.78 & 0.03 & 0.07)$\times$10$^{-5}$  \\ 
     190.6 &    218.8 &    204.0 &      2.6 &    0.288 &    0.007 &       5256 & ( 2.0 &  0.1)\% & (1.81 & 0.03 & 0.05)$\times$10$^{-5}$  \\ 
     218.8 &    251.2 &    234.2 &      2.9 &    0.290 &    0.007 &       4002 & ( 1.9 &  0.1)\% & (1.20 & 0.02 & 0.03)$\times$10$^{-5}$  \\ 
     251.2 &    288.4 &    268.9 &      3.4 &    0.291 &    0.007 &       2926 & ( 2.0 &  0.2)\% & (7.59 & 0.14 & 0.19)$\times$10$^{-6}$  \\ 
     288.4 &    331.1 &    308.8 &      3.9 &    0.291 &    0.007 &       2136 & ( 2.1 &  0.2)\% & (4.81 & 0.11 & 0.12)$\times$10$^{-6}$  \\ 
     331.1 &    380.2 &    354.5 &      4.4 &    0.290 &    0.007 &       1648 & ( 2.1 &  0.2)\% & (3.25 & 0.08 & 0.08)$\times$10$^{-6}$  \\ 
     380.2 &    436.5 &    407.1 &      5.1 &    0.292 &    0.007 &       1240 & ( 2.0 &  0.2)\% & (2.12 & 0.06 & 0.05)$\times$10$^{-6}$  \\ 
     436.5 &    501.2 &    467.4 &      5.8 &    0.291 &    0.007 &        889 & ( 2.2 &  0.2)\% & (1.32 & 0.05 & 0.03)$\times$10$^{-6}$  \\ 
     501.2 &    575.4 &    536.6 &      6.7 &    0.289 &    0.007 &        650 & ( 2.2 &  0.2)\% & (8.49 & 0.34 & 0.21)$\times$10$^{-7}$  \\ 
     575.4 &    660.7 &    616.1 &      7.7 &    0.288 &    0.007 &        536 & ( 2.0 &  0.2)\% & (6.13 & 0.27 & 0.15)$\times$10$^{-7}$  \\ 
     660.7 &    758.6 &    707.4 &      8.8 &    0.285 &    0.007 &        390 & ( 2.0 &  0.2)\% & (3.92 & 0.20 & 0.10)$\times$10$^{-7}$  \\ 
     758.6 &    871.0 &    812.2 &     10.2 &    0.284 &    0.007 &        271 & ( 2.3 &  0.3)\% & (2.38 & 0.15 & 0.06)$\times$10$^{-7}$  \\ 
     871.0 &   1000.0 &    932.5 &     11.7 &    0.278 &    0.008 &        195 & ( 2.3 &  0.3)\% & (1.52 & 0.11 & 0.04)$\times$10$^{-7}$  \\ 
    1000.0 &   1148.2 &   1070.7 &     13.4 &    0.276 &    0.008 &        136 & ( 2.6 &  0.4)\% & (9.29 & 0.82 & 0.27)$\times$10$^{-8}$  \\ 
    1148.2 &   1318.3 &   1229.3 &     15.4 &    0.274 &    0.009 &         74 & ( 3.6 &  0.5)\% & (4.38 & 0.53 & 0.14)$\times$10$^{-8}$  \\ 
    1318.3 &   1513.6 &   1411.4 &     17.6 &    0.267 &    0.009 &         93 & ( 2.2 &  0.4)\% & (4.99 & 0.53 & 0.17)$\times$10$^{-8}$  \\ 
    1513.6 &   1737.8 &   1620.5 &     20.3 &    0.263 &    0.010 &         33 & ( 5.0 &  0.9)\% & (1.52 & 0.28 & 0.06)$\times$10$^{-8}$  \\ 
    1737.8 &   1995.3 &   1860.6 &     23.3 &    0.255 &    0.011 &         26 & ( 5.4 &  0.9)\% & (1.07 & 0.22 & 0.05)$\times$10$^{-8}$  \\ 
    1995.3 &   2290.9 &   2136.3 &     26.7 &    0.249 &    0.012 &         17 & ( 5.8 &  0.9)\% & (6.24 & 1.61 & 0.30)$\times$10$^{-9}$  \\ 
    2290.9 &   2630.3 &   2452.8 &     30.7 &    0.243 &    0.014 &         12 & ( 7.9 &  1.1)\% & (3.84 & 1.20 & 0.21)$\times$10$^{-9}$  \\ 
    2630.3 &   3019.9 &   2816.1 &     35.2 &    0.233 &    0.015 &          4 & (18.2 &  2.5)\% & (1.03 & 0.63 & 0.07)$\times$10$^{-9}$  \\ 
    3019.9 &   3467.4 &   3233.4 &     40.4 &    0.227 &    0.017 &          4 & (15.4 &  2.4)\% & (9.53 & 5.64 & 0.70)$\times$10$^{-10}$  \\ 
    3467.4 &   3981.1 &   3712.4 &     46.4 &    0.218 &    0.018 &          4 & (11.2 &  2.6)\% & (9.07 & 5.12 & 0.77)$\times$10$^{-10}$  \\ 
    3981.1 &   4570.9 &   4262.4 &     53.3 &    0.210 &    0.020 &          3 & (11.4 &  4.0)\% & (6.15 & 4.02 & 0.60)$\times$10$^{-10}$  \\ 
\hline
\end{tabular}
\end{center}
\end{table}

\clearpage



\renewcommand{\refname}{References}

\newpage

\begin{figure}[!ht]
\includegraphics[width=1.0\textwidth]{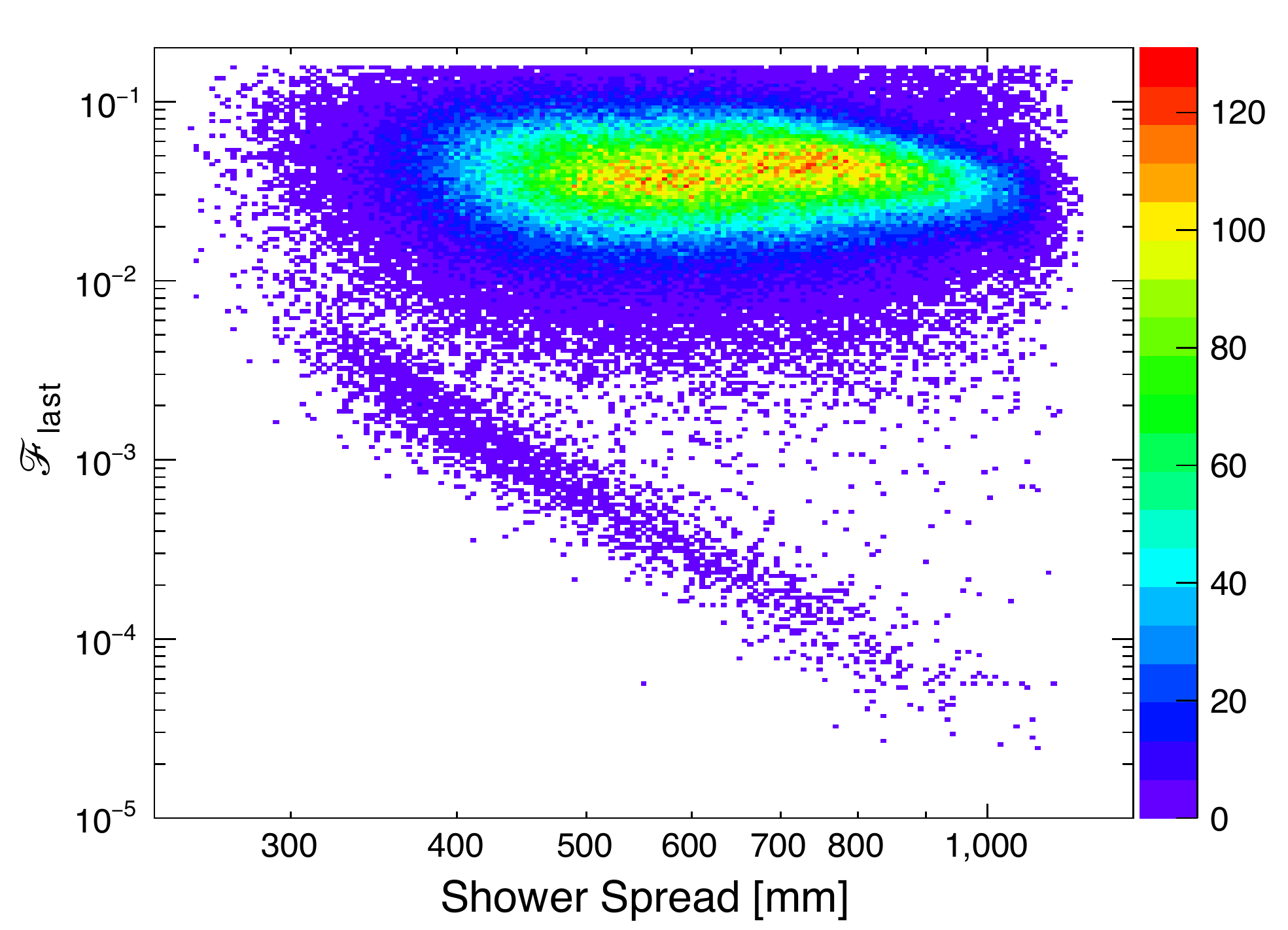}
\caption{{\bf Discrimination between electrons and protons in the BGO instrument of DAMPE.} Both the electron candidates (the lower population) and proton candidates (the upper population) are for the DAMPE flight data with deposit energies in the BGO calorimeter between $500~{\rm GeV}$ and $1~{\rm TeV}$.
${\cal F}_{\rm last}$ represents the ratio of energy deposited in the last BGO layer to the total energy deposited in the BGO calorimeter\cite{Chang2008a}. The shower spread is defined as the summation of the energy-weighted shower dispersion of each layer. The color palette on the right side represents the event number in each pixel.
}
\label{fig:e-p}
\end{figure}
\vfill

\begin{figure}[!ht]
\includegraphics[width=1.0\textwidth]{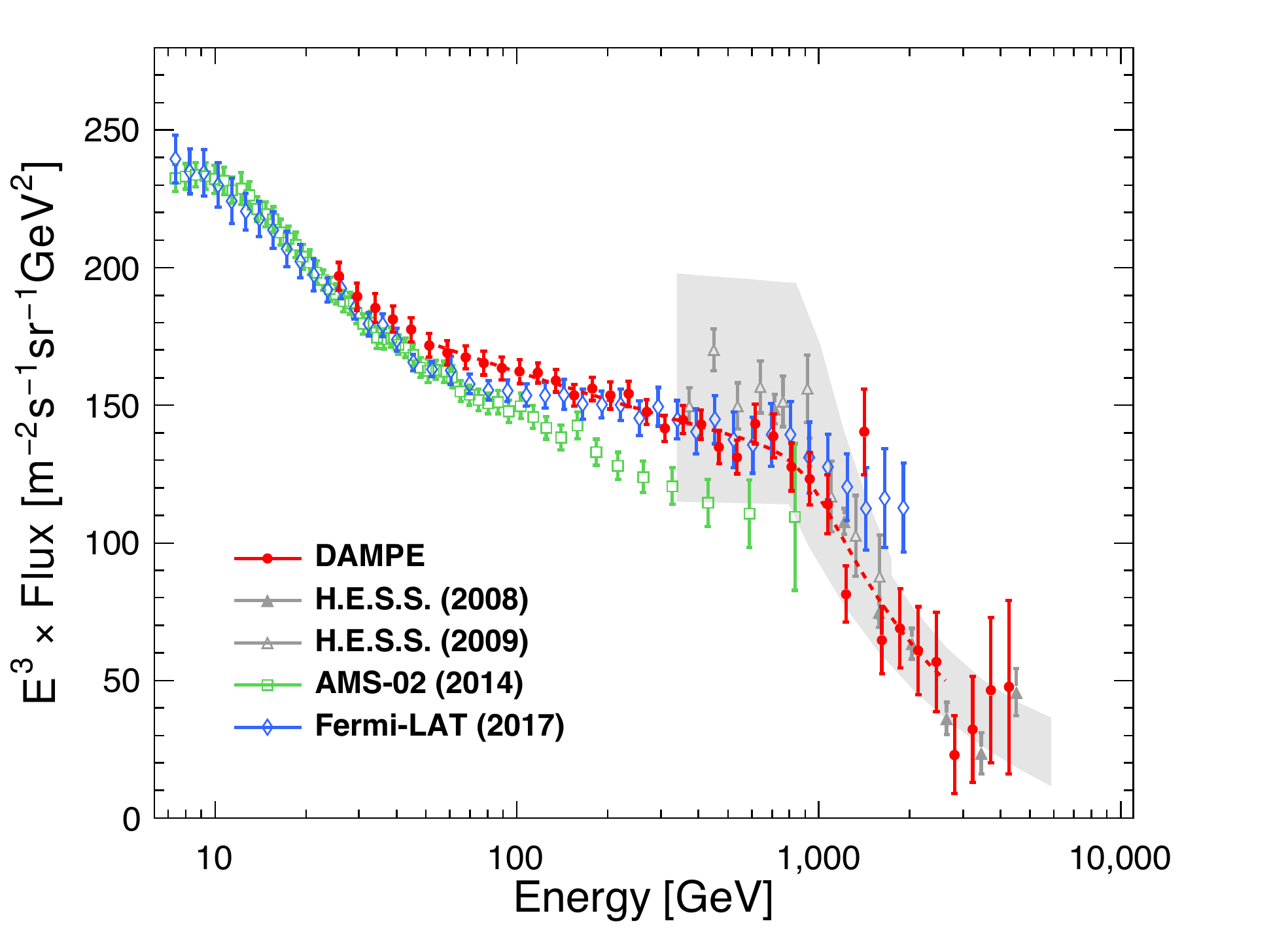}
\caption{{\bf The CRE spectrum (multiplied by $E^3$) measured by DAMPE.} The red dashed line represents a smoothly broken power-law model that best fits the DAMPE data in the $55~{\rm GeV}-2.63~{\rm TeV}$ range. Also shown are the direct measurements from space-borne experiments AMS-02\cite{Aguilar2014} and Fermi-LAT\cite{Abdollahi2017}, and the indirect measurement by H.E.S.S. (the grey band represents its systematic errors apart from the $\sim15\%$ energy scale uncertainty)\cite{Aharonian2008,Aharonian2009}. The error bars ($\pm$1 s.d.) of DAMPE, AMS-02 and Fermi-LAT include both systematic and statistical uncertainties added in quadrature.
}
\label{fig:spectrum}
\end{figure}
\vfill

\clearpage


\begin{addendum}

\item[Acknowledgements] The DAMPE mission is funded by the strategic priority science and technology projects in space science of Chinese Academy of Sciences. In China the data analysis was supported in  part by the National Key R\&D Program of China (No. 2016YFA0400200), National  Basic Research Program of China (No. 2013CB837000), NSFC (Nos. 11525313 and 11622327), and the 100 Talents Program of Chinese Academy of Sciences. In Europe the activities and the data analysis are supported by the Swiss National Science Foundation (SNSF), Switzerland; the National Institute for Nuclear Physics (INFN), Italy.

\item[Author Contributions] This work is the result of the contributions and efforts of all the participating institutes, under the leadership of Purple Mountain Observatory, Chinese Academy of Sciences. All authors have reviewed, discussed, and commented on the present results and on the manuscript. In line with collaboration policy, the authors are listed here alphabetically.


\item[Author Information] Reprints and permissions information is available at www.nature.com/reprints. The authors declare no competing financial interests.
Correspondence and requests for materials should be addressed to the DAMPE collaboration (dampe@pmo.ac.cn).

\end{addendum}

\newpage
\begin{center}
{\bf Methods}
\end{center}

{\bf Discrimination between electrons and protons.} The method of electron selection in this work relies on the difference in the developments of showers initiated by protons and electrons\cite{Chang1999,Schmidt1999,Chang2008a}. The procedure of the method is illustrated as follows. First, we search for events passing through the entire BGO calorimeter. We select events with hit positions from $-28.5$ cm to $28.5$ cm for the top layer and $-28$ cm to $28$ cm for the bottom layer (each BGO bar lays from $-30$ cm to $30$ cm). Second, we calculate the shower spread, expressed by the energy-weighted root-mean-square ($RMS$) value of hit positions in the calorimeter. The $RMS$ value of the $i$th layer is calculated as
\begin{equation}
{RMS}_i=\sqrt{\frac{\sum_j (x_{j,i}-x_{c,i})^2E_{j,i}}{\sum_j E_{j,i}}},
\end{equation}
where $x_{j,i}$ and $E_{j,i}$ are coordinates and deposit energy of the $j$th bar in the $i$th layer, and $x_{c,i}$ is the coordinate of the shower center of the $i$th layer. Figure~1 in the main text shows the deposited energy fraction in the last BGO layer (${\cal F}_{\rm last}$) versus the total $RMS$ value of all 14 BGO layers (i.e., $\sum_i{RMS}_i$). We can see that electrons are well separated from protons. Note that in Figure~1 and Extended Data Figure~1, heavy ions have already been effectively removed through the PSD selection based on the charge measurement.

Extensive Monte Carlo (MC) simulations are carried out to compare with data. Our MC simulations are based on {Geant4.10.02}\cite{geant4}. The hadronic model QGSP$_{-}$BERT is used to generate proton sample used in this analysis. We have compared the two hadronic models in Geant4, QGSP$_{-}$BERT and FTFP$_{-}$BERT, for the high energy range ($>50$ GeV), and found that the difference of the proton contamination estimate between the two models is less than $10\%$ for energies up to 5 TeV. The corresponding systematic uncertainty on the CRE spectrum measurement is thus negligible.

For a better evaluation of the electron/proton discrimination capabilities, we introduce a dimensionless variable, $\zeta$, defined as
\begin{equation}
\zeta={\cal F}_{\rm last}\times (\Sigma_i{RMS}_i/{\rm mm})^{4}/(8\times 10^{6}).
\end{equation}
The $\zeta$ distribution for events with deposited energies from 500 GeV to 1 TeV in the BGO calorimeter is shown in Extended Data Figure~1. Blue points represent the flight data, and red histogram represents the MC data (the electron MC data are in black and the proton MC data are in green). The MC data and the flight data are in good agreement with each other. A clear separation between electrons and protons is shown. The electron/proton discrimination capability with the $\zeta$ variable has also been validated with the 400 GeV proton data collected at the CERN beam test facilities using the DAMPE Engineering Qualification Model (EQM). The EQM is essentially the same as the final Flight Model, only except that 166 out of the 192 silicon ladders in STK are replaced by mechanical units, which have the same amount of material and mechanical/thermal properties as the real ones\cite{WuX2015}. In the data analysis, we take the cut of $\zeta=8.5$.

We also check the consistency of the $\zeta$ variable between the two-side readouts, the $P$ (positive) and $N$ (negative) sides, of the BGO crystal.
The gains of the two sides differ by a factor of $\sim5$. The data sets from each end can be used to measure the energy and hence provide the particle identification independently. The distribution of the ratio between the two groups of readout-based $\zeta$ values of CRE candidates agrees well with one another with no evidence of asymmetry between $P$- and $N$-sides, as illustrated in Extended Data Figure~2.

As independent analyses, the Principal Component Analysis (PCA)\cite{Jolliffe1986} and Boosted Decision Trees (BDT)\cite{Roe2005} classifier have been adopted for electron/proton discrimination, which give quite similar discrimination power as the {\it $\zeta$} (or equivalently, ${\cal F}_{\rm last}-\Sigma_i{\rm r.m.s}_i$) method. These three methods give well consistent (within the statistical uncertainties) results of the final CRE flux. In Figure~2 we present the $\zeta$ method-based spectrum.

{\bf Proton contamination estimate.} Based on the consistency between the MC simulations and the flight data (see Extended Data Figure~1), we use the simulation data as templates, which are normalized through fitting to the flight data, to estimate the proton contamination in the signal region. 
The contamination fraction is smaller than 6\% for energies below $\sim 2$ TeV without significant fluctuation. The estimated proton contamination has been subtracted in the final CRE spectrum shown in Figure~2 and the fluxes in Table 1.

{\bf Systematic uncertainties of the CRE flux measurement.} Systematic uncertainties of the flux measurement have been evaluated, with the dominant sources being related to the background subtraction and the effective acceptance (the product of the geometrical acceptance and the selection efficiency).

The systematic effect of the proton contamination estimate is evaluated by changing the modeling of $\zeta$ and the definition of the background region, taking into account also the limited Monte Carlo statistics at high energy. The results are reported in Table 1 (i.e., the background fraction).

The acceptance is determined by the BGO calorimeter. Since DAMPE has a precise tracking system, the reconstructed electron track can be used to define the incoming particle direction. The correlation between the track and the BGO shower direction is found to be well reproduced by Monte Carlo simulation. The residual difference was used to estimate the systematic uncertainty by varying the geometrical acceptance cut accordingly, resulting in a 2.2\% error which is independent of the particle energy.

For the selection efficiency, two components contribute the majority of the systematic uncertainties. One is the trigger efficiency. Its systematic uncertainty is evaluated by comparing the efficiency of the MC simulation to the one measured from the pre-scaled data sample collected with lower trigger thresholds (the unbiased and low energy triggers). The overall agreement is excellent and the difference is used to characterize the systematic uncertainty, with the level $1.5\%$ at 25 GeV and $1\%$ at 2 TeV, respectively. The other main systematic uncertainty arises from the $\zeta$ selection. After subtracting the small amount of proton candidates, the distribution of the $\zeta$ variable from the MC simulation is in a good agreement with the selected electron candidates from flight data. However, an energy-dependent difference between flight data and MC simulation is observed, which is also confirmed by the 250 GeV electron data taken at the CERN beam test. The MC $\zeta$ distribution is thus shifted to match the distribution of data, resulting in an efficiency correction of $-1.9\%$ at 25 GeV and $8.4\%$ at 2 TeV, respectively. The systematic uncertainty of the CRE spectrum due to this correction is estimated to be $1.0\%$ at 25 GeV and $4.2\%$ at 2 TeV.

The absolute energy scale uncertainty constitutes another type of systematic uncertainty that will shift the spectrum up or down coherently, without changing spectral features of the flux. For DAMPE, the absolute energy scale is estimated to be $\sim1.013$ times higher \cite{Zang2017}. Its small effect on the flux (i.e., scaled down by a factor of $\sim 2.6\%$) has not been corrected in this work.

{\bf Energy measurements.} The BGO calorimeter is a total absorption electromagnetic calorimeter. MC simulations show that the energy leakage from calorimeter bottom is negligible even at TeV energies because of the large thickness ($\sim 32$ radiation lengths). Despite the energy loss in the dead material, more than 90\% of the primary energy of an electron is deposited in the BGO crystals. An energy correction method taking into account the incident position, direction and the shower development is applied to each electron candidate\cite{YueChuan}. The beam-test data and the MC simulations show that the energy resolution of DAMPE is better than 1.2\% for electrons with energies from 100 GeV to 10 TeV\cite{Chang2017}.

In addition, we checked the consistency of the energy ratio of the CRE candidate measured from the $P$- and $N$-sides. Extended Data Figure~3 presents the ratios between the two sides together with a Gaussian fit which gives a mean of 1.005$\pm$0.005 and a sigma of 0.016$\pm$0.001, supporting the quoted $\sim1\%$ energy resolution from MC simulations\cite{Chang2017}.

{\bf Comparison of different spectral models.} We follow the procedures outlined in Appendix C of Abdollahi et al. (2017)\cite{Abdollahi2017} to fit the CRE spectrum in the energy range of $55~{\rm GeV}-2.63~{\rm TeV}$. The potential systematic uncertainties on the CRE flux measurement due to the $\zeta$ method (including the background contamination and the $\zeta$-selection)  is modeled by a set of nuisance correction parameters. The number of nuisance parameters is assumed to be ${\cal N}=6$, and the bin size was chosen with equal energy bin in logarithmic space. We fit the data with a single power-law model ($\Phi = \Phi_{0}(E/100~{\rm GeV})^{-\gamma}$) and a smoothly broken power-law model ($\Phi = \Phi_{0}(E/100~{\rm GeV})^{-\gamma_1}[1+(E/E_{\rm b})^{-(\gamma_{1}-\gamma_{2})/\Delta}]^{-\Delta}$ with the smoothness parameter $\Delta$ fixed to be 0.1\cite{Fermi2013}), respectively. The single power-law fit yields $\Phi =(1.64\pm0.01)\times 10^{-4}~(E/100~{\rm GeV})^{-3.13\pm0.01}$ ${\rm m^{-2}~s^{-1}~sr^{-1}~GeV^{-1}}$ with $\chi^{2}/{\rm d.o.f}=70.2/20$.
The broken power-law fit yields $\gamma_1=3.09\pm 0.01$, $\gamma_2=3.92\pm 0.20$, $\Phi_{0}=(1.62\pm0.01)\times 10^{-4}~{\rm m^{-2}~s^{-1}~sr^{-1}~GeV^{-1}}$, $E_{\rm b}=914\pm98$ GeV, and $\chi^2/{\rm d.o.f}=23.3/18$.
These fit results are shown in Extended Data Figure~4. Compared with the single power-law hypothesis, the $\chi^{2}$ value is smaller by 46.9 for two less degrees of freedom for the smoothly broken power-law hypothesis. The smoothly broken power-law model is thus strongly preferred (at the $6.6\sigma$ level).

{\bf Code availability.} { The numerical code has been developed with a dedicated application to the DAMPE data analysis. Due to the uniqueness of the DAMPE design and the complexity involved in the data analysis, the software package has limited application to the relevant community. We have opted not to make the code public.}

{\bf Data availability.} {The cosmic ray electron spectrum data, along with statistical and systematics uncertainties, are reported in Figure 2 and available in Table 1. The other data that support the plots within this paper and other findings
of this study are available from the DAMPE Collaboration (dampe@pmo.ac.cn) upon reasonable request.}

\renewcommand{\refname}{References}

\newpage

\renewcommand{\figurename}{Extended Data Figure}
\setcounter{figure}{0}

\begin{figure}[!ht]
\includegraphics[width=\textwidth]{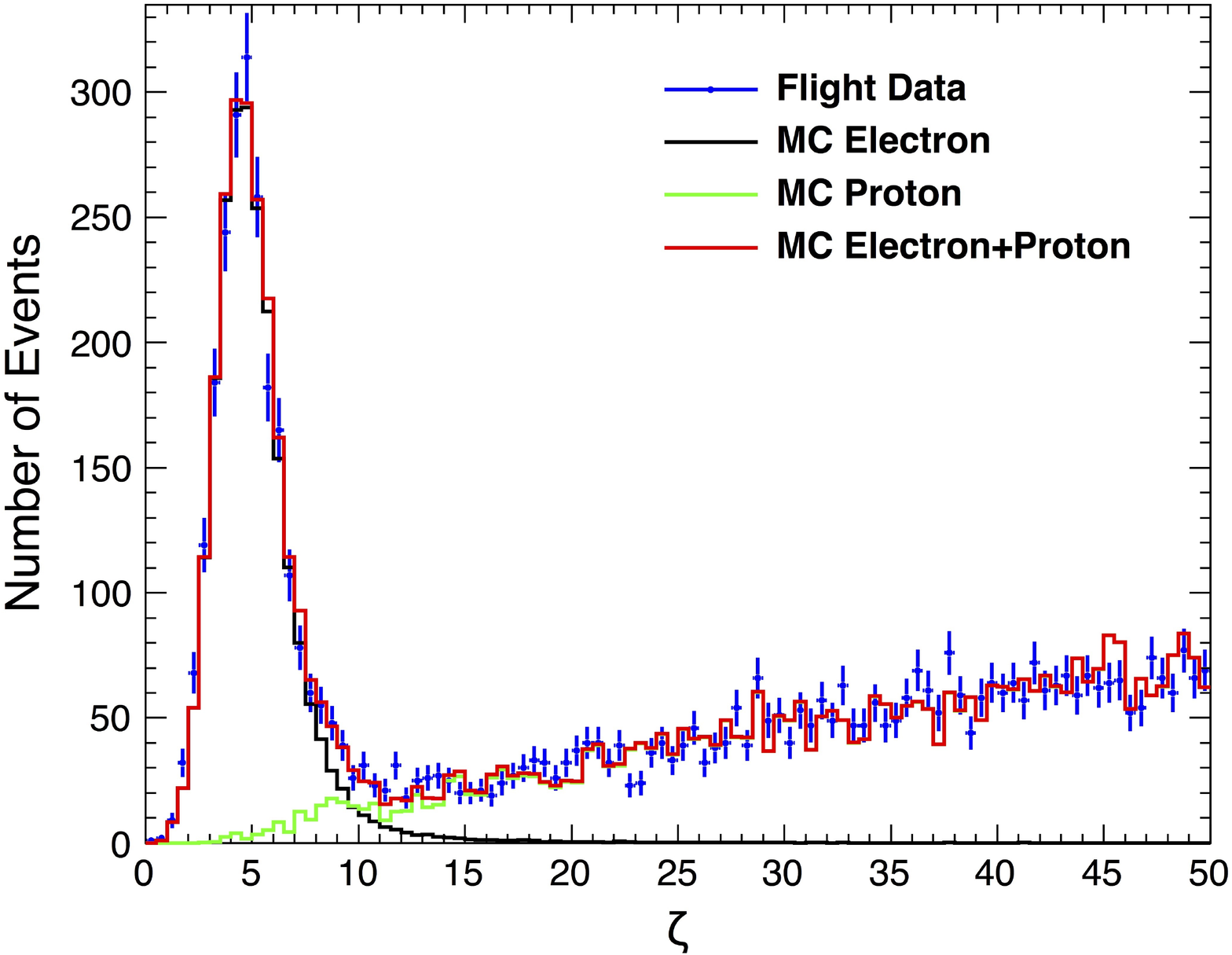}
\caption{{\bf Comparison of the flight data and the MC simulations of the $\zeta$ distributions}. All events have deposited energies between 500 GeV and 1 TeV in the BGO calorimeter.  The error bars ($\pm$1 s.d.) represent statistical uncertainties.  As for the MC simulation data, the black, green and red histograms represent the electrons, protons, and their sum, respectively.}
\label{fig:e-p-2}
\end{figure}
\vfill

\begin{figure}[!ht]
\includegraphics[width=1.0\textwidth,angle=0]{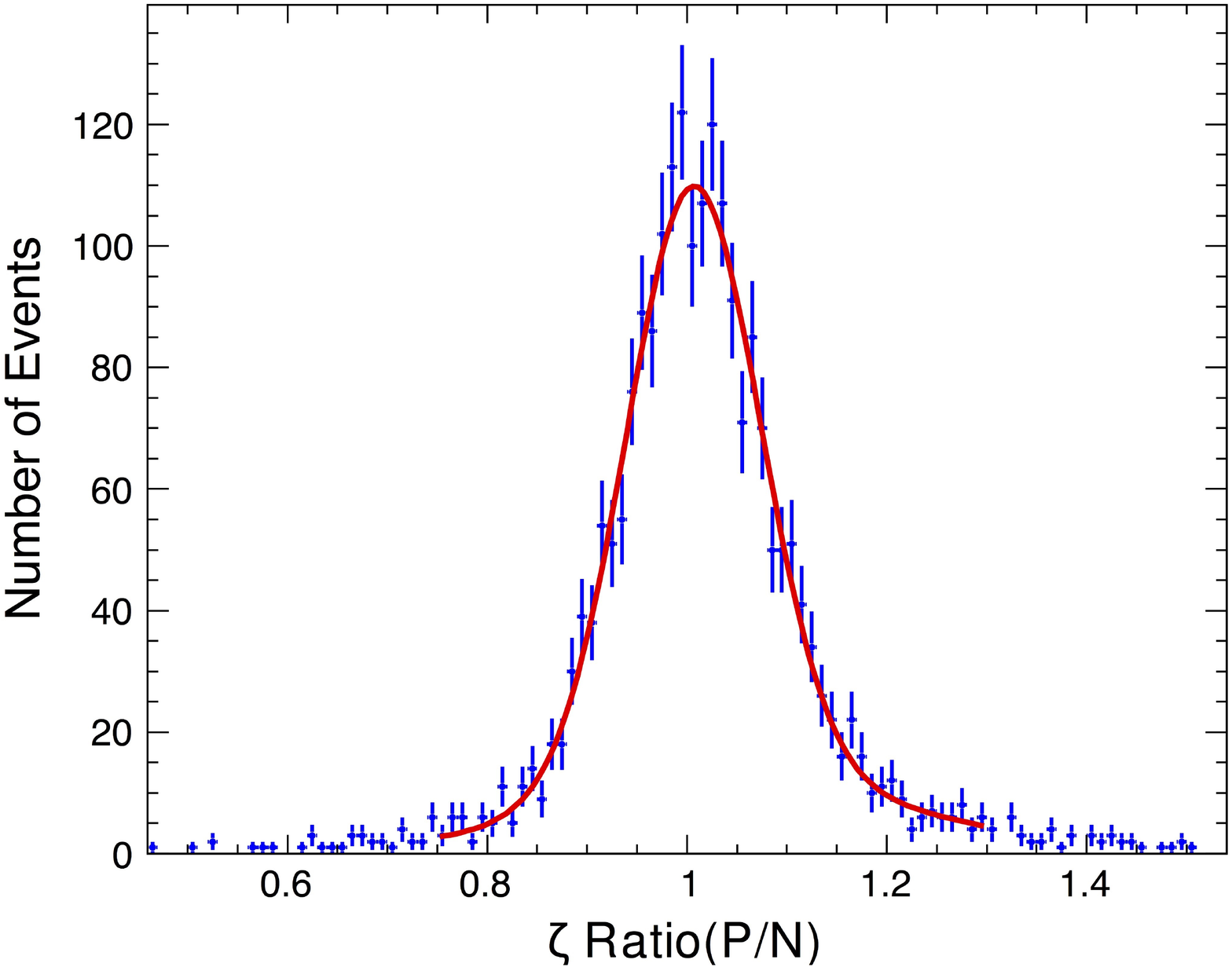}
\caption{{\bf Ratios of the $\zeta$ values calculated from the $P$- and $N$-side readout data.}
The events have deposit energies between 500 GeV and 1 TeV in the BGO calorimeter.
The error bars ($\pm$1 s.d.) represent statistical uncertainties. The red line represents a Gaussian fit to the data points. The mean of the ratios is $1.015\pm0.002$ and the width is $0.110\pm0.005$.}
\label{fig:N-P}
\end{figure}
\vfill

\begin{figure}[!ht]
\includegraphics[width=1.0\textwidth,angle=0]{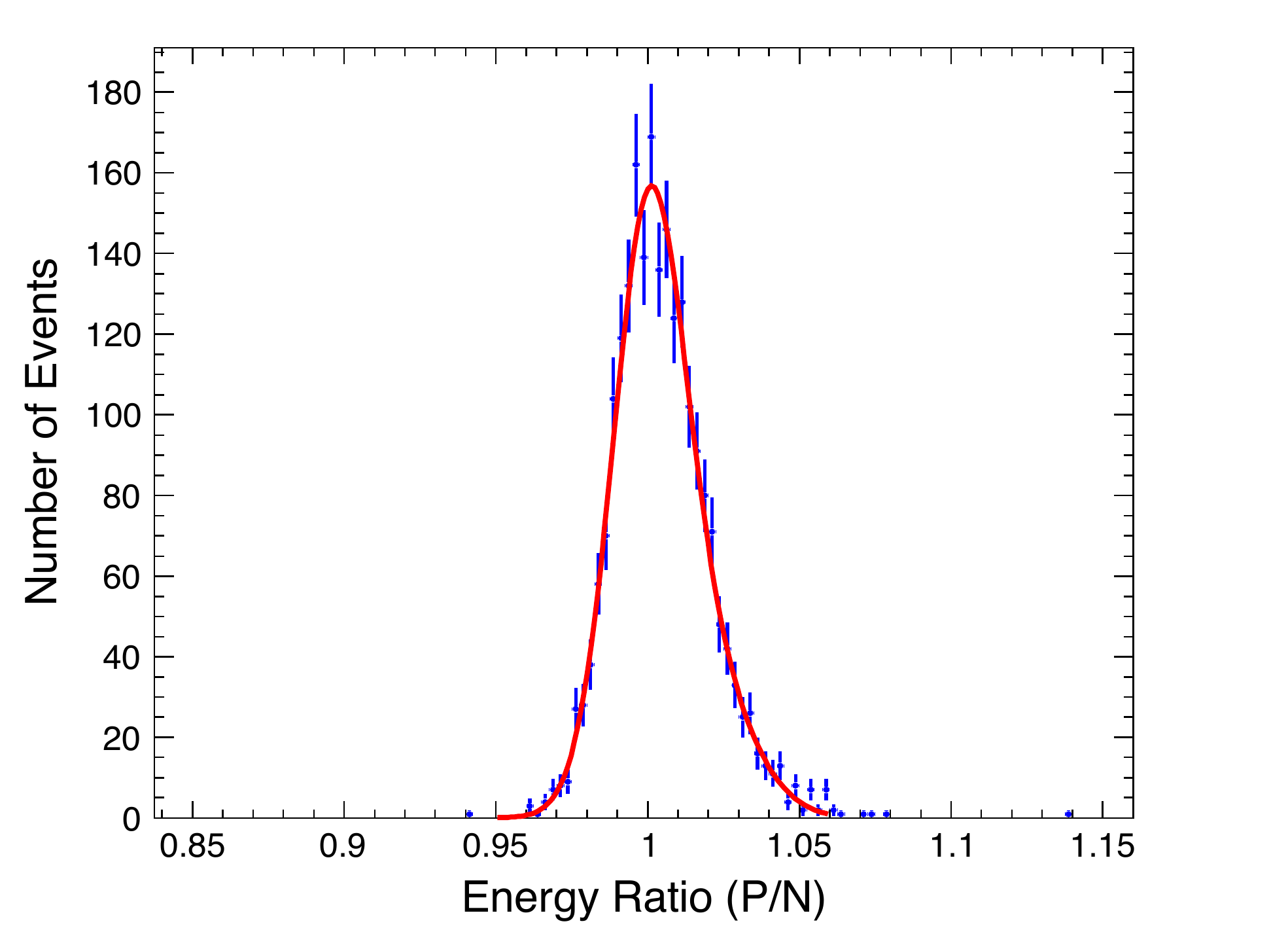}
\caption{{\bf Ratios of the energies reconstructed with the $P$- and $N$-side readout data}. All events have deposit energies between 500 GeV and 1 TeV in the BGO calorimeter. The error bars ($\pm$1 s.d.) represent statistical uncertainties. The red line represents a Gaussian fit to the data, with a mean of 1.005$\pm$0.005 and a sigma of 0.016$\pm$0.001.
}
\label{fig:PN-energy}
\end{figure}
\vfill

\begin{figure}[!ht]
\includegraphics[width=1.0\textwidth,angle=0]{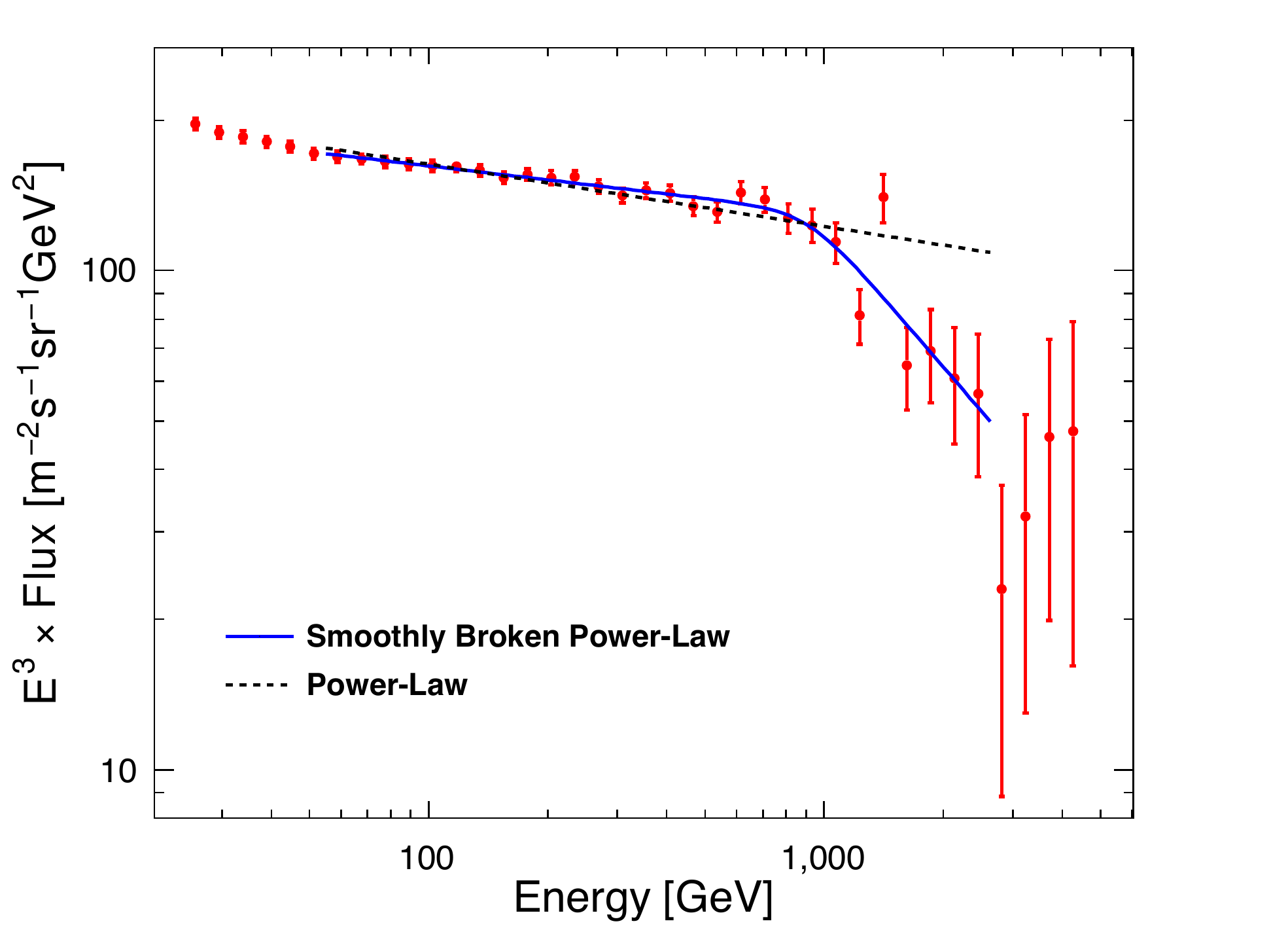}
\caption{{\bf Comparison of two spectral models for the DAMPE CRE spectrum.} The dashed and solid line show the best fit results of the single power-law and smoothly broken power-law models, respectively.}
\label{fig:residuals}
\end{figure}
\vfill

\clearpage


%
%

\end{document}